\documentstyle[epsfig,twoside]{l-aa}

\renewcommand{\epsfig}[1]{Missing author supplied figure}

\newcommand{\Msol}{M_\odot}
\newcommand{\Rsol}{R_\odot}

\begin{document}
\title{Experimental limits on the contribution of
sub-stellar and stellar objects to the galactic halo
\thanks{This work is based on observations made at the
European Southern Observatory, La Silla, Chile.}}
\author{EROS Collaboration}
   \author{
R. Ansari\inst{1}, F. Cavalier\inst{1}, 
M. Moniez\inst{1}, 
E. Aubourg\inst{2}, P. Bareyre\inst{2}, S. Br\'ehin\inst{2}, 
M. Gros\inst{2}, 
M. Lachi\`eze-Rey\inst{2}, B. Laurent\inst{2}, E. Lesquoy\inst{2}, C.
Magneville\inst{2},
A. Milsztajn\inst{2}, L. Moscoso\inst{2}, F. Queinnec\inst{2},
C. Renault\inst{2},
J. Rich\inst{2}, M. Spiro\inst{2}, L. Vigroux\inst{2},
S. Zylberajch\inst{2},  J.-P. Beaulieu\inst{3}, R. Ferlet\inst{3}, Ph.
Grison\inst{3},
A. Vidal-Madjar\inst{3}, 
J. Guibert\inst{4}, O. Moreau\inst{4}, F. Tajahmady\inst{4}, E.
Maurice\inst{5},
 L. Pr\'ev\^ot\inst{5}, C. Gry\inst{6},
  }
\institute{
Laboratoire de l'Acc\'el\'erateur Lin\'eaire, Universit\'e de Paris-sud, IN2P3,
Centre d'Orsay, F-91405 Orsay Cedex, France
\and
CEA, DSM, DAPNIA, Centre d'Etudes de Saclay, F-91191 Gif-Sur-Yvette, France
\and
Institut d'Astrophysique de Paris, 98 bis Boulevard Arago, F-75014 Paris,
France
\and
Centre d'Analyse des Images de l'INSU, Observatoire de Paris, 61 avenue de
l'Observatoire, F-75014 Paris, France
\and
Observatoire de Marseille, 2 place Le Verrier, F-13248 Marseille Cedex 04,
France
\and
Laboratoire d'Astronomie Spatiale de Marseille, Traverse du Siphon, Les Trois
Lucs, F-13120 Marseille, France
}

   \offprints{M. Moniez; {\it see also our WWW server
at  URL} \ {\tt http://www.lal.in2p3.fr/EROS/eros.html}}
   \date{Received aa/bb/95, accepted xx/yy/95}

\maketitle
  \markboth{R. Ansari et al. : Experimental limits on the contribution of
sub-stellar and stellar objects to the galactic halo}{R. Ansari et al. :
Experimental limits on the contribution of
sub-stellar objects to the galactic halo}

\begin{abstract}
EROS (Exp\'erience de Recherche d'Objets Sombres) has been monitoring
the luminosity of 4 million stars in the Large Magellanic Cloud
in order to search for gravitational microlensing by unseen
objects in the galactic halo. We present here the results from
3 years of EROS Schmidt plates data.
Two stars exhibit light curves that are consistent with a
sizeable microlensing effect.
CCD data obtained later on revealed
that one of these stars is an eclipsing binary system.
Combining Schmidt plates data and the published results
from our 16 CCD camera, we
set upper limits on the number of unseen objects in the halo
in the mass range $[10^{-7},1]\Msol$.

\end{abstract}
\section{Introduction}
The study of rotation curves of spiral galaxies has
led to the hypothesis of the presence of
massive extended halos of ``dark matter" (e.g. \cite{primac}).
Numerous candidates have been proposed as
constituents of this matter, such as weakly interacting
massive particles (WIMPS), massive neutrinos, or ``brown dwarf"
stars, i.e. stellar bodies lighter than the thermonuclear ignition
limit ($M<0.07\Msol$ to $0.1\Msol$, depending on
the metallicity) (\cite{carr}).
Their existence could more specifically solve the
question of the missing {\em baryonic} mass, which emerged
from primordial nucleosynthesis studies (e.g. \cite{Kolb}).

Paczy\'nski (\cite{pacz}) pointed out the possibility of using
the gravitational microlensing effect to detect such massive
halo objects.
Four experimental teams, EROS (\cite{erpre},b, 1995),
MACHO (\cite{machlmc}, 1995a,b),
OGLE (\cite{oglpr}, 1994a) and DUO (\cite{duo}) are now
searching for those effects on resolved light sources,
EROS and MACHO having started
with the search for the effect on LMC stars.

A halo object passing close enough to the
line of sight of a star in the Large Magellanic Cloud (LMC),
would temporarily magnify its light (microlensing effect).
At a given time $t$ the
apparent light amplification is determined by
\begin{equation}
A(t)=\frac{u(t)^2+2}{u(t)\sqrt{u(t)^2+4}}\ ,
\end{equation}
where $u(t)$ is the distance between the undeflected
line of sight and the deflecting object, expressed
in units of the ``Einstein Radius"
\begin{eqnarray}
R_E & = & \sqrt{\frac{4GM}{c^2}\ Lx(1-x)} \\
\ & \simeq & 970\times\sqrt{\frac{M}{\Msol}}\times\sqrt{\frac{L}{10\ Kpc}}
\times\frac{\sqrt{x(1-x)}}{0.5}\times\Rsol\ . \nonumber
\end{eqnarray}
Here $G$ is the gravitational constant,
$L$ is the distance to the source, $xL$ is the distance to
the deflector and $M$ its mass.
The motion of the deflector relative to the line of sight
to the source makes the magnification vary with time:
for deflectors of masses within the interval
$[10^{-7},1]\Msol$ located in the halo, time scales range
typically from hours to months for a significant variation
in the magnification of a source in the LMC.
Assuming a deflector moving at a constant apparent transverse
speed $V_T$, reaching its minimum
distance to the undeflected line of sight (impact parameter) $u_{min}$
at time $t_0$, $u(t)$ is given by
$u(t)=\sqrt{u_{min}^2+((t-t_0)/\Delta t)^2}$.
%

The ``lensing time scale" $\Delta t=\frac{R_E}{V_T}$ is
the only measurable parameter
bringing useful information about the deflector.
For a source in the LMC ($L = 55 kpc$),
it can be expressed as:
\begin{equation}
\Delta t (days)=91\times
\left[\frac{V_T}{200 km/s}\right]^{-1}
\times \left[\frac{M}{\Msol}\right]^{\frac{1}{2}}
\times\frac{[x(1-x)]^{\frac{1}{2}}}{0.5}
\end{equation}
The probability for a given star of the LMC
to be amplified by a factor larger than 1.34 ($u=1$) at a given time,
is the probability for its line of sight to intercept the
Einstein disk of radius $R_E$ of one of the deflectors.
This probability, the optical depth $\tau$, scales
with the total mass of the halo. Details of the halo geometry fix
the value of the proportionality coefficient.

The standard isothermal halo model we use assumes a total galactic mass of
$4.10^{11}\Msol$ within 50 kpc of the galactic centre, and a radial
density $\rho(r)$ decreasing as :
\begin{equation}
\rho(r)=\rho_0\times\frac{a^2}{r^2+a^2}
\end{equation}
where $r$ is the distance to the galactic centre. We
have taken $a=7.8$ kpc for the core radius (\cite{Caldw}).
For this model, the optical depth $\tau$ is about $4.5\times10^{-7}$ up to
the LMC.
The velocity distribution of the deflectors is assumed to be
boltzmannian, with a dispersion of 245 km/s.
Assuming that all deflectors have the same mass M,
the rate per star for microlensing effects with amplifications
greater than a threshold amplification $A_T$ (corresponding to an
impact parameter $u_{min}=u_T$) has been calculated to be
$1.6\ 10^{-6}\ u_T\sqrt{\Msol/M} \ yr^{-1}$
(\cite{Griest}).

The microlensing effect has some very characteristic
features which should enable to discriminate it from any known
intrinsic stellar variability :
\begin{itemize}
\item The event is singular in the history of the source
(as well as of the deflector), during the experiment life time.
\item The gravitational origin of the effect implies that the
magnification is independent of the colour.
\item The luminosity is a known function of
time, depending on only 3 parameters ($u_{min}, t_0, \Delta t$),
with a symmetrical shape.
\end{itemize}
The latter two characteristics are affected if there is a deviation
from the approximations of a single
point-like deflector and source and of uniform motion (see section 6).
In addition, statistical tests may be applied to a set of
microlensed stars :
\begin{itemize}
\item As the geometric configuration of the source-deflector system
is random, the impact parameters
of the events have to be uniformly distributed.
This allows the prediction of the theoretical
amplification distribution which, corrected for
microlensing detection efficiency,
can be compared with the experimental one.
\item The passive role of the lensed stars implies that their
population should be representative of the monitored
sample, particularly with respect to colour and magnitude.
\end{itemize}
\section{The observations}
To fully cover the range of possible microlensing time scales,
EROS has performed two observing programs, one using a
16 CCD camera mounted on a 40 cm diameter telescope (\cite{ccdcam},
\cite{erccd}),
to search for short time scale microlensing phenomena,
and the other using Schmidt photographic plates. We present
here results from the latter program, superseding our previous
publication (\cite{erlmc}).

During 3 annual periods of about 6 months, 290 usable photographic
plates (in total) of $29\times29$ cm$^2$ have been exposed at the E.S.O
Schmidt telescope (1 meter free aperture, f/3).
Half of the
plates (098-04 emulsion) were taken with a RG630 red filter
and half (IIaO emulsion) with a GG385 blue filter.
Exposure times were 1 hour in each colour, and apart from the
very crowded LMC bar region,
our star detection efficiency abruptly drops at a limiting
magnitude of around 20.5 in red and 21.5 in blue.
The data taking period is limited by the maximum
excursion of the telescope around the meridian position
($\pm2.5$ hours).
The time sampling of the plates (see figure \ref{fg1}),
\begin{figure}
\begin{center}
\mbox{\epsfig{file=fig/date.eps,width=8.8cm}}
\caption[]{\it Time distribution of the photographic plates taken
for EROS  (from 1990 to 1993).
\label{fg1}}
\end{center}
\end{figure}
makes the experiment sensitive to
microlensing event durations ranging from a few days to a few months.
Two thirds of the plates were taken during the second year, and the
first period was irregularly sampled.
The seeing, comprised in the interval [0.9, 2.] arcsec has an average value of
about 1.5 arcsec.
The scale of the images is 67.5 arcsec/mm and
the usable field covered with a plate is $5.2^{\circ}\times5.2^{\circ}$,
centered
on position $(\alpha = 5h20mn,\delta = -68^{\circ}30')$ (eq. 2000).
The transmission coefficient T of each plate was digitized to 0.8
giga-pixels (12 bits) of 10 microns size (0.675 arcsec) using the
``MAMA"\footnote{MAMA is developed and operated by INSU/CNRS.}
microdensitometer
(Machine Automatique \`a Mesurer pour l'Astronomie) at the
Observatoire de Paris (\cite{Berger}). The digitization takes 6 hours per
plate.
\section{Data reduction}
We have divided the field into 28x28 sub-fields of 1cm$^2$ each,
and have checked that the response of a plate digitised by the
``MAMA" measuring machine
does not significantly vary at the centimeter scale.
Each stage of the following analysis is then
performed locally for each sub-field.
A home-made photometric program has been developed, optimised
for the search for variations in crowded fields, which have an
occupation rate of about 100 reconstructed stars/arcmin$^2$.
A detailed description of the analysis can be found in
(\cite{Aubourg}, \cite{Cav}, \cite{Laur}).
\subsection{From pixels to light curves}
The quantity $\Phi_T=[(1-T)/T]^{0.6}$ varies
approximately linearly with the flux $\Phi$ collected on the
photographic plate in the magnitude range $[17,21]$ (blue)
$[16,20]$ (red).
We therefore use $\Phi_T$ in our photometric fitting and starfinding
procedures. The thus obtained star magnitudes are then
corrected using an empirical formula providing
an accuracy of 0.2 magnitude in average,
which is derived from comparison of
selected fields with CCD
measurements taken with the E.S.O-Danish 1.54m telescope.
As a check of the validity of this empirical formula over the
whole plate field,
we have measured that
the mean magnitude and magnitude dispersion of the red
giant clump are constant within 0.07 mag
for each 1cm$^2$ sub-field where those numbers can be
correctly estimated,
i.e. outside the LMC bar.

The first step of the reduction was
to construct a reference catalogue of stars for each colour,
using a composite image obtained by adding 10 aligned and resampled
images, taken under good conditions. From such a composite
image, the detection of stars is
improved due to the better signal/noise ratio.
Their positions
are measured with a
better precision than on a single image, allowing to use these as
reference positions. Their luminosities are also used as reference ones.
We have detected 8 million stars in at least
one colour, while 6 million are seen in the two colours.

Each 1 cm$^2$ sub-field of each digitized plate is then submitted
to the following treatment, leading to new luminosity measurements
for each catalogued star :
\begin{itemize}
\item The geometrical correspondence between a current image
and the reference catalogue is determined
with a positioning precision better than $2.5 \mu m$ (0.17 arcsec)
using a pattern recognition program, which associates the brightest
stars detected by a crude algorithm, with the brightest stars of
the reference catalogue.
\item The point spread function
(PSF) of a star centered at position $\vec{r_0}$
is described by a Moffat Function (\cite{Mof})
$\Phi(\vec{r})=A\times\
(1+\frac{\mid\vec{r}-\vec{r_0}\mid^2}{\sigma^2})^{-3}$.
{}From a sample of well defined and well isolated stars, we determine
the parameter $\sigma$ of this PSF, approximately
independent of the flux of the stars (except for those which
saturate the plate, or are underexposed).
\item Using the geometrical transformation,
the position of each catalogued star is predicted
on the current image, and the luminosity of the
stars are measured by fitting only one parameter per star (the
amplitude A of the PSF, centered on the predicted
position).
This procedure is considerably faster, and also
more precise, than fitting simultaneously the positions {\em and}
the fluxes of each star.
To extract the flux of the stars in a crowded environment,
the amplitudes of all
the stars and a common background within a $0.5\times0.5$mm$^2$
sliding window with 0.3mm step are fitted
together, and only the measurements of the stars in the
$0.3\times0.3$mm$^2$ central part are retained.
\item The incoming fluxes on the plates depend on
observational conditions such as atmospheric
transmission, seeing or exact exposure time.
In particular, measured luminosities are affected by
the seeing in a way that
depends on the environment of the stars, due to
the imperfect knowledge of the PSF and to the
limits of validity of the procedure.
We correct for the differences in the data taking conditions
by requiring that the mean luminosity of stars in a given
luminosity interval and for a given estimated
background due to neighbouring stars
be equal to their mean reference luminosity.
\end{itemize}
A file, one for each colour, containing the time series of
luminosity measurements (hereafter light curves)
is then updated with these corrected values for each star.
The uncertainty $\sigma_i$ on a luminosity measurement $\Phi_i$
depends mainly on the luminosity of the star
and on the global quality of image number $i$, determined
by the seeing and the sky background.
For each 1cm$^2$ sub-field of each plate i, we determine
empirically the resolution $\sigma_i$ as a function of the luminosity,
by comparing the star fluxes $\Phi_i$ with their reference
values to estimate standard deviations. A detailed description of
the parametrisation of $\sigma_i$ can be found with our
atlas of LMC stars (\cite{ercat}).

The performance of this reconstruction procedure is summarized in
figure \ref{fg2}, which shows the mean dispersion of the
measurements along the light curves as a function of the
magnitude $m_B$ (and $m_R$) in the EROS blue (red) filter band,
for stars having at least 90 reliable measurements for each colour.
\begin{figure}
\begin{center}
\mbox{\epsfig{file=fig/reso.eps,width=8.8cm}}
\caption[]{\it Relative plate to plate average dispersion of the luminosity
measurements versus magnitude
(upper= EROS blue, lower= EROS red).
This dispersion is taken as an estimator of the photometric precision.
The superimposed hatched histograms show the magnitude distribution of
the stars.
\label{fg2}}
\end{center}
\end{figure}
\subsection{Analysis of the light curves}
The next step of the analysis is to test each light curve for
the presence of a microlensing event. Algorithms to be applied
to the light curves to search for microlensing candidates
should accept light curves exhibiting one and only one
positive fluctuation that is greater than that expected
from measurement errors. Additionally, the magnifications
in the two colours should be equal within errors and the
temporal development of the magnification consistent with
the one expected from a microlensing effect.

Two independent analyses using different approaches have been
developed in order to select the microlensing candidates.
The first one was intended to extract a clear and
isolated signal, with a high signal/background ratio
appropriate to our ignorance of the risk of false events.
It uses criteria based on the expected shape
of a microlensing effect, which are relevant only for
sufficiently sampled events.
The other analysis includes the
search for undersampled events,
with a higher detection efficiency for events of a few days.
It does not take into account the shape of the light variation
and uses criteria mainly based on fluctuation probabilities.
The efficiency
of each of the following cuts to accept real microlensing
events has been studied and optimised with
Monte-Carlo generated lensing light curves (section 5).
\subsubsection{First analysis of the light curves}
This analysis (\cite{Cav}) starts from a sample of 4.2 million stars
with reliable measurements in both colours.
Given this large number of light curves, we
define a reduced sample using a crude selection filter to remove
stable stars which did not exhibit any significant fluctuation.
We adjust the filtering in order to select about 10\% of the light curves.

Each light curve from this prefiltered set (0.44 million stars)
is then subjected to
a series of tests to select microlensing events:
\begin{itemize}
\item
We first require the existence of a bump (i.e. at least
3 consecutive measurements above the base flux by more than
$1.2\sigma$) in both colours, each having its
maximum flux within the duration of the other.
\item
Then we compute the correlation coefficient of the blue and red
luminosity variations during the bumps, and require this coefficient
to be larger than a given threshold, such as to reject
the 40\% objects with the most chromatic fluctuations.
\item
In order to have well defined base fluxes for stability studies,
we limit the duration of the bump to a maximum of 100 days.
At this stage of the analysis, 7140 stars are selected.
\item
Then we fit the experimental light curves with
a theoretical microlensing function defined for
each colour $\lambda$ as the product of the amplification
function $A(t)$ by a constant base flux $F_{\lambda}$.
Parameters $u_{min}$ (linked to the maximum amplification
$A_{max}$ through eq. (1)), $t_0$, $\Delta t$ and $F_{\lambda}$ are
adjusted independently for each colour using least-squares fits.
A global fit is also performed on the whole set of measurements,
constraining the function A(t) to be the same for the two colours.

This fitting procedure
provides reliable fitted parameters $t_0$ and $\Delta t$
for 707 pairs of light curves, the other ones mostly showing
erratic variations.
\end{itemize}
Figure \ref{fg3}a shows the distribution of the red magnitudes for
these 707 stars. This distribution is dominated by the
brightest stars, amongst which the fraction of variable stars is high.
\begin{figure}
\begin{center}
\mbox{\epsfig{file=fig/fig3.eps,width=8.8cm,height=10.cm}}
\caption[]{\it Magnitude $m_R$ of selected stars at various stages
of the first analysis
\\
(a) : before applying the last two sets of cuts.
The grey histogram shows the
shape of the distribution for the unbiased initial sample of 4.2 million stars.
\\
(b) : after requiring the shape and achromaticity criteria.
\\
(c) : after requiring the stability of the star outside the bump.
\\
The two marked events are the only ones surviving all the cuts.
\label{fg3}}
\end{center}
\end{figure}

Two sets of cuts concerning independent measurements are finally used :
\begin{itemize}
\item
The first one combines a test of shape and a loose achromaticity
requirement using only the measurements taken
{\em during} the bump.
To test for the shape, we require the $\chi^2$ of
the global fit mentioned above to be less than 2.5/d.o.f.
For testing achromaticity, we
require the $\chi^2$ obtained by fitting
the red measurements imposing the $A(t)$ function adjusted in blue,
to be less than 4/d.o.f. (loose cut).
206 stars survive this set of cuts. If we consider that this
sample is still dominated by a background of classical types of stars
(stable or variable),
the rejection power of the set is $707/206 \simeq 3.4$.
\item
The second set of cuts tests the stability of the light curve
{\em outside} the bump.
We first require the correlation between the blue and red light
curves to be weak outside this bump, and therefore compatible
with random measurement errors.
We also require the calculated probability of
the most important fluctuation with respect to a constant flux
found outside the main bump
to be more than $10^{-9}$ for both colours.
13 stars out of 707 survive this set of cuts, which then
has a rejection power of $707/13 \simeq 54$.
\end{itemize}

Figure \ref{fg3}b (c) shows the distribution of the red magnitudes for
stars which satisfy the first (second) set of cuts.
Only the 2 stars with the magnitudes marked with an arrow on
figure \ref{fg3} satisfy both sets of cuts.
\subsubsection{Background}
The 707 stars with bumpy structures are mainly variable
stars lying within the instability strip and the red giant branch.
Except for the two selected stars, they exhibit either obviously
periodic variabilities or a main chromatic or asymmetrical
bump with other smaller variations.
A few may also be
intrinsically stable stars with groups of deviating measurements.
For this mixture of stars, the global rejection of the
last two sets of cuts should be larger than the product
$3.4\times54=186$, because they are anti-correlated:
if a significant bumpy structure is found, then the probability
for the corresponding star to be a true variable having
other significant variations outside this bump is higher
than for stars which do not satisfy the first set of cuts.
In the sub-sample of stars with R magnitude larger than 17
we find that less than 0.13 star
from the above mixture are expected to accidentally survive
the selection process, and
be considered as microlensing-like events.
In this statistical calculation, only the variable types
effectively present in the sample of 707 stars
can be accounted for, and it is not
possible to estimate the contamination due to some other rare
type of variable star.
Anyway, we should consider
such
contamination as a ``signal" in the sense that, using
our data only, there is no way to distinguish it from a real
microlensing effect.
\subsubsection{Second analysis of the light curves}
This analysis, described in detail in (\cite{Laur}), is restricted to the
3.33 million best measured stars, located at least 3~mm away from the
outer edges of the
analysed region (the $28 \times 28$ 1cm$^2$ sub-fields),
well isolated and with magnitudes within or close to the linear regime,
and whose light curves are correctly sampled in both colours
during all three seasons.
%
Twenty five plates showing abnormally large seeing or response
dispersions or non-circular star images are excluded from the analysis.
Each light curve is then subjected to the following selection
process.

A search for the first and second most significant fluctuations
is done for each colour, using the probability of the $\chi^2$
of successive measurements with respect to the base flux
(the base flux is defined as the average flux in the two seasons
where the star luminosity is lowest).
Each selected fluctuation should contain at least 3 measurements
more than 0.5~$\sigma$ from the base flux and one measurement
more than 2.0~$\sigma$ from the base flux.
In addition, the
first fluctuation is required to correspond to a luminosity increase.
We do not allow fluctuations to span more than one season
(see figure \ref{fg1}).
\begin{itemize}
\item
We require that light curves exhibit significant first
fluctuations in both colours, with a modest (at least 8~\%)
time overlap, leaving 6912 stars.
\item
We then require the second fluctuation to have both a low
absolute significance and a low significance relative to
the first one. The aim here is to reject most periodic
variable stars, except those with periods larger than a year or
smaller than a few days.
\item
In the remaining sample of only 44 stars, 34 exhibit long
time scale fluctuations (at least two thirds of a season)
with slow luminosity variation. These stars, most of which
are bright, are removed by requiring that the dispersion
of the luminosity measurements be significantly higher
during the season of the first (main) fluctuation than during
the other two seasons.  This cut rejects long-period
variable stars.
\item
We then require that the correlation coefficient between the
blue and red light curves outside the main fluctuations
be compatible with zero at the 4 standard deviation level.
This cut rejects three short period variable stars.
\item
{}From the 7 remaining stars, only 4 pass
a loose achromaticity test (compatibility of
the variations for the 2 colours within 5 standard deviations).
\item
We found that 2 of those 4 stars are located 4 arcsec apart
and are contaminated by a nova (IAU circular 5651).
They passed the selection process
because there is no criterion based on the shape
of the variations.
The other 2 light curves are the same as found in the
first analysis.
\end{itemize}
The fact that the two different selection processes
isolate {\em the same two light curves} reinforces their status as
peculiar events, whatever their final interpretation.
\section{Complementary observations}
We started detailed observations on the two candidates
to further check their validity as possible microlensed stars.

Old plates have been examined to check the constancy of
the fluxes on a time scale of a few years.
High resolution imagery has been taken at the E.S.O New
Technology Telescope. We found no pathologic feature
(like the presence of bright nearby stars).
No anomaly appears from UBVRI photometry of the two candidates.
The two stars have been found to lie within the LMC from
spectral measurements (\cite{erspec}).
They are also from two distinct stellar types (see table 1):
EROS \#1 candidate
is a B star with H emission lines, that may indicate the presence
of surrounding gas,
and EROS \#2 is an A star.

To test for any further variability,
photometric measurements have been made using our
16 CCD camera, mounted on a 40 cm diameter telescope (\cite{ccdcam}).
Some photometric measurements were also provided
by the 70 cm Geneva telescope at E.S.O. in La Silla.
Data concerning the candidates have also been extracted
from a series of plates exposed in 1994 for our experiment.
Finally, the EROS \#2 candidate has profited by a specific
treatment which allowed us to recover a total of 142
measurements from the 1990-93 plates ;
as the star lies close to the edge of the plates,
its neighbourhood is truncated and despite the fact that the star
is present on each plate, the pattern recognition phase of the
geometrical alignment procedure, which needs the neighbouring
stars, had failed in a large number of cases.

{}From the new CCD measurements, we discovered that EROS \#2 is an
eclipsing binary star with period $2.8169\pm\ 0.0005$ days,
and amplitude at the 0.5 magnitude level (\cite{variab}).
More measurements are needed to elucidate the nature of
the singular high amplitude bump we detected.
\section{Calculation of the microlensing detection efficiency}
To determine the efficiency of each cut,
we have applied them to Monte-Carlo generated
light curves of microlensing events.
We have developed two independent simulation programs of
the microlensing effect, using different ways to take into
account the imperfectly known photometric uncertainties.

The first program generates light curves through the
following procedure:
the base fluxes of a lensed star in the two colours are
randomly chosen following our experimental
magnitude versus colour distribution of catalogued stars.
Then we calculate the lensing effect due to the crossing of
an object of fixed mass ($10^{-5}$ to $1\Msol$),
at impact parameter $u_{min}$ in the interval
$[0, 2]$, with distance to the Earth and speed
spanning the standard halo model (\cite{Caldw}).
The time of maximum amplification $t_0$ was picked at
random in a 3 year period containing all photographic plates
of figure \ref{fg1}.
Then a light curve is generated, with
our experimental time sampling, and with random gaussian
shifts corresponding to the uncertainties measured from
the data for such luminosities.

The other program starts from our experimental light curves, and
superimposes the effect of an amplification from a microlensing
on the measured values, to produce new light curves
and error bars. Microlensing light curves were simulated
with
time scales $\Delta t$ in the range $[0.3, 600]$ days,
and with the same $u_{min}$ and $t_0$
parameters distributions than in the first program.

The light curves produced by the two simulation programs have been
submitted to the selection programs,
to provide us with an expected number of events in the hypothesis
of a halo made of equal mass objects. For a given analysis, the
two programs give compatible estimates.

Figure \ref{fg4} shows the detection efficiency for the second analysis,
normalized to the microlensing events with impact parameter
$u_{min}\le 1$,
\begin{figure}
\begin{center}
\mbox{\epsfig{file=fig/efficacite.eps,width=8.8cm}}
\caption[]{\it EROS Schmidt plates microlensing detection efficiency
$\epsilon(\Delta t)$
as a function of event time scale $\Delta t$, from our second analysis.
\\
$\epsilon(\Delta t)$ is the ratio of
the number of detected events with any $u_{min}$
to the number of events generated over a full 3 year period,
with $u_{min}\le 1$.
\label{fg4}}
\end{center}
\end{figure}
as a function of the time scale $\Delta t$ of the microlensing effect.
Figure \ref{fg7}a (curve labelled ``Plates")
shows the expected number of events in the second
analysis as a function of
the mass of the deflectors, assuming a delta distribution for
the halo object masses.
%
\section{Study of systematic uncertainties}
\subsection{Uncertainties from the simulation programs}
The two simulation programs give compatible estimates within 5\%
when used to calculate the detection efficiency of a given selection program.
The difference can be explained by the different generations
of the experimental errors.
This 5\% is taken as the systematic uncertainty for
the detection efficiency that is associated with our imperfect
knowledge of the experimental resolution.

The simulation programs also use the
generated amplifications as measured amplifications
thus assuming that the plates are correctly calibrated.
The experimental magnitudes have been empirically derived by
comparison with CCD images (see section 3.1),
but the slope of the calibration line cannot be obtained
with a precision better than 5 percent.
This induces a 5 percent uncertainty in the expected number of
events.
\subsection{The single source approximation}
The efficiency calculations described above assume that
the light curves represent the light from only one star so that
the entire flux would be microlensed in an event.
In fact, a large proportion of stars are in binary
systems and
the microlensing light curve may be distorted
because of different amplifications for the two components.
However, this would affect our detection efficiency
only if the two stars are of comparable magnitude
and have a projected separation in the deflector's plane
larger than half the Einstein radius.
The fraction of such pairs is sufficiently small that the effect
on our efficiency calculation should be negligible (\cite{GriestHu}).

On the other hand,
in crowded fields one must consider the possibility
that stars are accidentally ``blended'' so that a light curve
receives significant contributions from two or more
stars whose images overlap.
This effect can be studied by reconstructing artificially
produced images of known star content.
Previous studies have shown an efficiency reduction of $\sim 20\%$
for the MACHO collaboration survey of the central
regions of the LMC (\cite{Machofirstyear})
and   $\sim 10\%$ for the EROS CCD survey of the LMC bar (\cite{erccd}).
We expect a smaller effect for the present study because
of the smaller density of stars studied
($1.2\; 10^{5}$ resolved stars per square degree)
compared to MACHO
($9.5\; 10^{5}$ stars per square degree)
and EROS CCD
($1.9\; 10^{5}$ stars per square degree).
Additionally, the fact that we reconstruct
most of the red giant clump stars means that there are fewer stars just
below detection threshold compared to the EROS CCD study
where the threshold was near the middle of the clump.

Studies with Monte Carlo fabricated images (\cite{Laur})
have confirmed that this effect of blended stars is small,
contributing an error
of less than 10 percent to the expected number of events.
We include this as part of our systematic error.
\subsection{The point-like source approximation}
The efficiency calculations also assume that the sources are
point-like objects.
This approximation breaks down when the size of the star projected
onto the plane of the deflector is comparable to the impact
parameter.
For the deflector masses studied here, this can only affect
very high amplification events and has no effect on our detection
efficiency.
\subsection{The single lens approximation}
A small fraction of the deflectors are expected to be multiple
systems giving more complex microlensing light curves ;
a few percent of the observed microlensing light curves do display
such features (\cite{ogldl}, \cite{duodl}).
Anyhow the second analysis, which do not
require a specific shape for the light curves, is able to
detect them, except in the rare cases where two peaks are
clearly separated. The detection efficiency of our experiment,
calculated under the assumption of single lenses, should then not be
significantly affected.
\subsection{Loss of efficiency from subsequent requirements}
The use of the subsequent CCD observations to test the
two candidates for more criteria (see section 4) could in principle
reduce the detection efficiency of the experiment.

We estimate that less than 1\% of the monitored stars
do not lie in the LMC, and must be substracted from the
number of sources used to calculate our detection efficiency.
This can be considered as a negligible effect.

We have also checked that in the fields specially monitored with
the camera, a negligible fraction of stars do exhibit fluctuations
larger or equivalent to the fluctuations of EROS \#2.
This means  that if we could systematically perform
the same complementary observations that were done
for the two candidates, and remove sources with
variability at a level comparable to EROS \#2,
we would reject a negligible fraction of microlensing
targets.
\subsection{Uncertainties from the Halo model}
We have investigated the sensitivity of the expected number of
 microlensing events to the values of the parameters in the Halo
 model, keeping the total mass within 50~kpc at
 $4. 10^{11} \Msol$.
 The core radius has been varied between 3 and 8~kpc ;
 the distance of the Sun to the Galactic centre has been varied
 by 0.5~kpc ; the average velocity of halo objects was varied by
 10~\% ; we have allowed for the uncertainty in the LMC velocity.
 Adding all the resulting changes in number of events in
 quadrature, one finds a typical error from these parameter
 uncertainties of 15~\%.
 Flattening the Halo along the Galactic pole axis by up to a
 factor of three decreases the expected number of events by
 at most 10~\%.
 We have also used the Halo model parameters of (Griest 1991) ;
 they correspond to expected number of microlensing events
 typically 15~\% higher than those given here (figure \ref{fg7}a).
\section{Discussion}
\subsection{Microlensing fits and statistical features}
Figures \ref{fg5} and \ref{fg6} show the light curves of the two
selected stars with
the fitted microlensing curves, on a truncated time scale.
The fit of a microlensing effect is superimposed onto the modulation
found with our complementary observations for the EROS \#2 candidate
(see \cite{variab} for details).
\begin{figure*}
\begin{center}
\mbox{\epsfig{file=fig/fig_cand1.eps,width=14cm}}
\caption[]{\it Light curves of the EROS \#1 candidate.
Data recorded with the blue (red) filter are presented in the
upper (lower) figure. One should note the non-continuous horizontal
axes : the three parts correspond respectively to data recorded in
1991, 1992 and 1993. Microlensing curves calculated with the global fitting
procedure of the first analysis are superimposed on both light curves.
\label{fg5}}
\end{center}
\end{figure*}
\begin{figure*}
\begin{center}
\mbox{\epsfig{file=fig/fig_cand2.eps,width=14cm}}
\caption[]{\it Light curves of the EROS \#2 candidate.
On the lower part, the fit of the
function of a microlensing variation combined with
the periodic variability of the eclipsing system is
superimposed on the 1991 data points.
\label{fg6}}
\end{center}
\end{figure*}
Separate fits were first calculated
for both colours and were found compatible within errors,
allowing us to use combined fits. This feature of
the light curves, together with the good $\chi^2$ values,
make them compatible with a microlensing effect.
Table 1 gives the relevant fitted parameters for the two candidates.
\begin{table}
\caption[]{Characteristics of the two microlensing candidates}
\begin{flushleft}
\begin{tabular}{lll}
\hline\noalign{\smallskip}
& EROS \#1  & EROS \#2\\
\noalign{\smallskip}
\hline\noalign{\smallskip}
Coordinates of 	& $\alpha=5h26m34.1s$	& $\alpha=5h06m05.2s$	\\
star (eq. 2000) & $\delta=-70^{\circ} 57'45"$	& $\delta=-65^{\circ} 58'33"$\\
$m_B$		& $19.4\pm 0.2$		& $19.2\pm 0.2$		\\
$m_R$		& $18.9\pm 0.2$		& $19.1\pm 0.2$		\\
Type of star	& B6-7 Ve or IVe	& A0-2 V		\\
Radial velocity	& $350\pm 170 km/s$	& $300\pm 155 km/s$	\\
Date of maximum	& 1 Feb. 1992		& 29 Dec. 1990		\\
amplification							\\
Event time scale	& $23\pm 2$		& $29\pm 2$		\\
$R_E/V_T$ (in days)						\\
Max. amplification	& $1.0\pm 0.1$		& $1.1\pm 0.1$		\\
(in magnitudes)							\\
Impact parameter	& $0.44\pm 0.02$		& $0.40\pm 0.03$		\\
(in Einstein radius)							\\
$\chi^2$ of combined fit	& 131/279 d.o.f.		& 154/273 d.o.f.		\\
							\\
Remarks		&			& binary system		\\
		&			& period 2.8 days		\\
\noalign{\smallskip}
\hline
\end{tabular}
\end{flushleft}
\end{table}

The two candidates have spatial locations and magnitudes
which are typical in our monitored sample.
The maximum magnifications we have observed
would correspond to lens configurations with reasonable
probabilities (impact parameters respectively 0.44 and 0.40
Einstein radius).

Given the unusual natures of our candidates,
to be conservative, we have for now to consider them either as
possible unusual types of variable stars or as microlensing events.
This means that when establishing statistical upper limits on the
standard halo mass from our observations, we consider
that we have observed 2 or less microlensing events.
\subsection{Constraints on the galactic halo}
We discuss now consequences of our observations in an
increasing order of model-dependence.

\subsubsection{Signal}
Taking into account the maximum statistical contamination
of 0.13 event expected from the known types of variable stars
(section 3.2.2.), we find
that the probability to find one event or more from those known origins
is 12\%, and the probability to find two or more events is only 0.8\%.

We have estimated the expected number of microlensing events
from deflecting stars
located in the LMC itself to be less than 0.2 in our experiment,
and the number from stars in the galactic disk to be less than 0.15.
One should notice that the fact that our candidates lie far
from the LMC bar makes even more unlikely
the possibility of lensing by LMC stars.
The probability to find one or more events from the known types
of variable stars, or from a microlensing effect due to LMC or galactic
disk stars is 38\%, and the probability to find two or more events
is 8.4\%.

\subsubsection{Optical depth and event rate}
The optical depth, given by the fraction of time
during which a star undergoes a lensing amplification
larger than 1.34, is computed from
$$\tau =\frac{\pi}{2.E}\sum_{events}
\frac{\Delta t}{\epsilon (\Delta t )}$$
where $E=3.33\ 10^6 stars \times 3\ years=1.0\ 10^7 star \times year$
is our exposure and $\epsilon (\Delta t )$ our detection efficiency as a
function
of $\Delta t$, normalized to the lensing events with $u_{min}\le1$ (fig.
\ref{fg4}).
If we assume that our two candidates are due to real microlensing effects,
then the corresponding detected optical depth is
$8.2\ 10^{-8}$.

The event rate is computed to be
$\Gamma = 1/E\times\Sigma_{events}(1/\epsilon(\Delta t))
\ =\ 7.3\ 10^{-7} events/star/year$.
\subsubsection{Masses of the deflectors}
For given time scales $\Delta t$ of 23 and 29 days,
we have computed for all possible masses the probability to
pick a microlensing event with duration $\Delta t$, or a
less probable one. Mass intervals where this probability is
higher than 5\% are $[0.01-0.7]\Msol$ (resp. $[0.02-1.1]\Msol$)
for $\Delta t=23$ (resp. 29) days.
\subsubsection{Limits on the composition of the galactic halo}
Figure \ref{fg7}a, established using the Monte-Carlo simulation, shows
the number of events expected to be selected by the second analysis,
for the standard halo described in section 1,
assuming that all lenses have the same mass.
This distribution is plotted for the presently described EROS Schmidt
plates data and also for the EROS CCD experiment (\cite{erccd}),
the results of which are almost completely
independent because of the different sets of monitored
stars and time samplings.
Adding in quadrature all the experimental systematic uncertainties discussed
in sections 6.1 to 6.5, one gets a 15\% global systematic uncertainty on
the number of events expected in the EROS Schmidt plates experiment.
As discussed in section 6.6, excursions of the halo parameters
compatible with the standard model contribute an additional
15\% uncertainty on the expected number of events.
\begin{figure}
\begin{center}
\mbox{\epsfig{file=fig/expect.eps,width=8.8cm}}
\caption[]{\it
\\
(a) : expected number of events assuming all
deflectors have the mass given in abscissa, in
the Schmidt plates data, in the CCD data, and in total.
\\
(b) : the excluded contribution at 95\% C.L. of massive objects
to the standard spherical halo, expressed in mass within 50 kpc (left scale)
and in standard halo fraction (right scale), assuming a delta
distribution for the mass of the deflectors. The excluded region at 95\%
C.L. from the MACHO experiment is also drawn (\cite{machlt}, b).
\label{fg7}}
\end{center}
\end{figure}

Figure \ref{fg7}b
gives the maximum contribution of massive objects to the standard
halo compatible with our observations at 95\% C.L.,
as a function of their mass
(obtained for the case of delta function mass distributions).
We establish this exclusion contour making use of the combined
sensitivities of the two EROS experiments and taking into account
the time scales of the two detected events (23 and 29 days)
in a simple and conservative way ;
from the estimated number and $\Delta t$ distribution of
the events expected with given mass deflectors, we compute the
fraction of the standard halo for which the probability to find
0, 1 or 2 events in the $[13,52]$ days time scale interval, multiplied
by the probability to find 0 event outside this interval,
is lower than 5\%.

{}From the two EROS experiments, we conclude that at 95\% C.L.,
objects of masses in the interval $[10^{-7},10^{-1}]\Msol$ cannot
contribute more than 50\% to the mass of the standard halo,
and objects of masses in the interval $[3.10^{-5},10^{-2}]\Msol$ cannot
contribute more than 25\%.

The most probable halo fraction
and deflector mass published by the MACHO collaboration (\cite{machlt})
are obviously not excluded by our results.

The results presented here are to a large extent independent of
the results of the MACHO collaboration (\cite{machlt}) also shown on
fig \ref{fg7}b, because 80\% of the EROS measurements were
made before MACHO data taking and the sets of monitored stars have
a small intersection, due to the chosen
fields and different limiting magnitudes.

We stress that our sensitivity is not large enough to put
constraints on the galactic dark matter distribution if it
exhibits a geometrical shape
significantly different from the standard halo model.
This would be for example the case of a thick disk or a
strongly flattened halo (more than a factor of 8).
\section{Conclusion}
We have searched for microlensing events with durations
ranging from a few days to a few months.
Using two independent analysis, we found two events
which could be interpreted as
microlensing events due to objects
with probable masses between $10^{-2}$ and $1\Msol$ if they lie in the
galactic halo.
One of the two selected stars was later found to be an eclipsing
binary system, and cannot be considered as a reliable candidate.
It is anyway not possible to rule out the possibility of
having detected new types of irregular variable stars,
and we expect to improve our
knowledge about the two candidates from various observations.
Combining the results presented here from the Schmidt plates experiment,
and earlier ones from our CCD experiment, allows us
to put significant constraints on the contribution
of massive compact objects to the mass of the standard galactic halo.
\begin{acknowledgements}
We thank A. Bijaoui for discussions,
G. and O. Pizarro for taking the photographic material, and
the MAMA team, particularly R. Chesnel and P. Toupet for their
participation to plate scanning.
\end{acknowledgements}

\end{document}